\documentclass{Interspeech}

\interspeechcameraready

\title{Spotlight-TTS: Spotlighting the Style via Voiced-Aware Style Extraction and Style Direction Adjustment for Expressive Text-to-Speech}

\author[]{Nam-Gyu}{Kim}
\author[]{Deok-Hyeon}{Cho}
\author[]{Seung-Bin}{Kim}
\author[affiliation={\dagger}]{Seong-Whan}{Lee}

\affiliation[nocounter]{Department of Artificial Intelligence}{Korea University}{Seoul, Korea}
\email{ng\_kim@korea.ac.kr\thanks{\hspace{-0.5em}$^\dagger$Corresponding author}, dh\_cho@korea.ac.kr, sb-kim@korea.ac.kr, sw.lee@korea.ac.kr}

\usepackage{comment}
\usepackage[utf8]{inputenc}
\usepackage{amsmath}  
\usepackage{amssymb}
\usepackage{pifont}
\usepackage{multirow}   
\usepackage{booktabs}   
\usepackage{graphicx}   
\begin{document}
\maketitle
\setcounter{footnote}{0}

\begin{abstract}
Recent advances in expressive text-to-speech (TTS) have introduced diverse methods based on style embedding extracted from reference speech. However, synthesizing high-quality expressive speech remains challenging. We propose Spotlight-TTS, which exclusively emphasizes style via voiced-aware style extraction and style direction adjustment. Voiced-aware style extraction focuses on voiced regions highly related to style while maintaining continuity across different speech regions to improve expressiveness. We adjust the direction of the extracted style for optimal integration into the TTS model, which improves speech quality. Experimental results demonstrate that Spotlight-TTS achieves superior performance compared to baseline models in terms of expressiveness, overall speech quality, and style transfer capability. Our audio samples are publicly available.\footnote{{https:/Spotlight-TTS.github.io/}}
\end{abstract}

\section{Introduction}
Text-to-speech (TTS) \cite{ren2021fastspeech} aims to synthesize speech from input text. With recent advancements in deep learning technology \cite{557671,602073,LEE1995783}, the naturalness of synthesized speech has improved significantly \cite{NEURIPS2022_69c754f5,cho24_interspeech}. Despite the development of general TTS systems, synthesizing human-like speech for applications such as virtual assistants and audiobooks remains challenging due to the lack of expressiveness of synthesized speech. To address this limitation, expressive TTS with style modeling and transfer techniques is attracting more attention. In particular, style transfer TTS systems are becoming increasingly important for real-world applications, as they eliminate the need for style-annotated datasets or matched pairs of text and speech data.  

Well-designed style encoder is a key component for achieving natural style transfer in TTS systems. The early approaches utilized sentence-level style by applying pooling operations \cite{pmlr-v80-skerry-ryan18a,pmlr-v80-wang18h,pmlr-v139-min21b}. However, the speaking style cannot be fully expressed by sentence-level style alone, as it has limitations in representing temporal variations in speech, which led to the syllable-level intonation modeling to better capture time-varying style \cite{tang2023qi}. Furthermore, GenerSpeech \cite{NEURIPS2022_4730d10b} explored a more fine-grained approach by extracting the frame-level style. This approach employs multi-level style encoder with vector quantization variational autoencoder \cite{NIPS2017_7a98af17} to encode style into a discrete codebook, serving as a bottleneck to eliminate non-style information. This approach leads to improvements in style transfer. However, there remains room for further improvement.

Recent works \cite{seong24b_interspeech, zhang-etal-2024-tcsinger} utilized advanced vector quantization methods to improve style extraction such as residual vector quantization (RVQ) and clustering style encoder (CSE). Despite these advances in style extraction methodology, several fundamental challenges remain unsolved. They treat all temporal segments equally, failing to capture the varying importance of different speech regions in extracting speaking style. Additionally, the straight-through estimator \cite{bengio2013estimating} used during backpropagation disregards the relative positioning of encoded features within each codebook region, limiting the model’s ability to learn fine-grained style details. Finally, relying solely on quantization as a bottleneck without additional constraints does not ensure that extracted style embeddings are independent of content, increasing the risk of content leakage during style transfer. Addressing these issues requires methods that enhance expressiveness while maintaining strong disentanglement from content.

In this paper, we address these limitations through voiced-aware style extraction and style direction adjustment. The importance of region-specific processing has been recognized in signal compression \cite{jayant1993signal}, and has recently gained renewed attention in deep learning \cite{huang2023not,Liu_Gui_Luo_2023}. Similarly, different regions of speech have varying impacts on speaking style. Voiced regions, which contain rich acoustic information such as harmonics generated by vocal fold vibration, are particularly important for style characteristics. In this context, our voiced-aware style extraction enables effective quantization based on a more compact representation, which enhances the preservation of style-related features with finer granularity. Additionally, we replace the conventional straight-through estimator with a rotation trick \cite{fifty2025restructuring} in our quantization process, which enables more precise style extraction, particularly in voiced regions. Furthermore, we introduce style direction adjustment, which adjusts the extracted style by modifying its angle using content and prosody vectors in the embedding space. This adjustment process effectively removes content information from style and prevents training instabilities caused by disentanglement. 
\begin{figure*}[!t] 
    \centering
\includegraphics[width=0.98\linewidth]{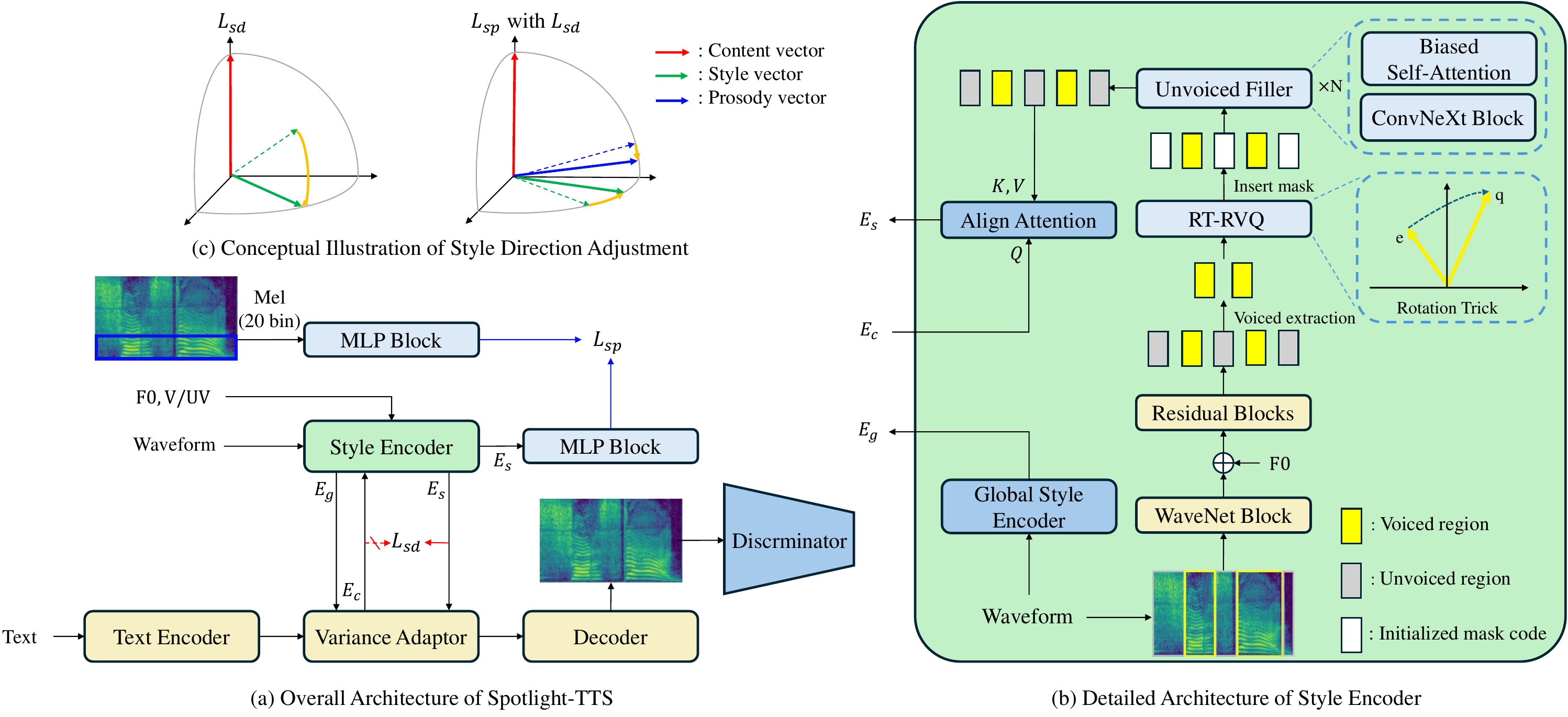}\vspace{-0.3cm}
\caption{(a) shows the overall architecture of our proposed Spotlight-TTS. \( E_{\text{c}}\), \( E_{\text{g}}\), and \( E_{\text{s}}\) denote the content embedding, global style embedding, and style embedding respectively. (b) shows the details of the style encoder. e and q represent the input feature of the quantization layer and quantized vector respectively. (c) shows the conceptual illustration of angles of vectors changed by style direction adjustment. \( L_{\text{sd}}\) and \( L_{\text{sp}} \) represent the style disentanglement loss and style preserving loss.}
\label{model}\vspace{-0.3cm}
\end{figure*}
Experimental results demonstrate that the proposed model not only achieves more expressive style transfer but also significantly improves the disentanglement of content and style, leading to more natural speech quality and pronunciation.  

\section{Spotlight-TTS}
In this section, we introduce our proposed model Spotlight-TTS. As shown in Figure \ref{model}, we focus on effectively extracting style by considering the importance of different Mel-spectrogram regions while adjusting the direction of style. Our proposed method consists of two parts: voiced-aware style extraction and style direction adjustment.

\subsection{Voiced-aware style extraction}
In speech synthesis, we hypothesized that style extraction through codebook learning could be improved by considering different Mel-spectrogram regions that contribute unequally to speaking style. As shown in Figure \ref{model} (b), voiced regions consist of harmonics that are highly correlated with speaking style, and unvoiced regions have simple repetitive patterns that are less related to style. Therefore, we propose voiced-aware style extraction, a novel style extraction method that focuses on the voiced region and fills in the unvoiced region with the unvoiced filler (UF) module.

\subsubsection{Voiced frame processing}
We use an RVQ module \cite{lee2022autoregressive} to extract detailed style embeddings. Given that unvoiced regions are less relevant to style compared to voiced regions, we focus RVQ processing on voiced frames through voiced extraction (VE). We use pre-extracted voiced and unvoiced (V/UV) flags to aggregate only voiced frames from the intermediate features as input to the RVQ module. During quantization, we adopt the rotation trick (RT) \cite{fifty2025restructuring} to improve gradient flow through the quantization layer. Unlike the conventional straight-through estimator, RT preserves the angle between the loss gradient and codebook vector during backpropagation. The RT is computed as:
\begin{equation}
\tilde{q} = sg\left[\frac{\left\Vert q\right\Vert}{\left\Vert e\right\Vert}R\right] e,
\end{equation}
where \(R\) is the rotation transformation that aligns the input feature \(e\) to the closest codebook vector \(q\) and \(\frac{\left\Vert q\right\Vert}{\left\Vert e\right\Vert}\) rescales \(e\) to match the magnitude of \(q\). Here, $sg \left [  \cdot  \right ]$ denotes the stop-gradient operator used to detach rotation and scaling terms from gradient computation.
During forward pass, we use \(\tilde{q}\), the transformed input feature, instead of the quantized vector. \(\tilde{q}\) is identical to \(q\), in the forward pass, ensuring the RVQ output remains unchanged. By rewriting \(q\) in terms of \(e\), the backward pass rotates the gradient to capture the relative position of \(e\) in the codebook.

\subsubsection{Unvoiced filler module}

To improve style continuity between different regions, we propose a UF module. After quantization, we generate learnable mask code embeddings with uniform random initialization and insert them into the unvoiced positions based on their positional information. The UF module then fills these embeddings with meaningful acoustic information. The UF module consists of $N$ identical sub-modules, each of which consists of a ConvNeXt block \cite{liu2022convnet} and a biased self-attention. Biased self-attention enables information flow from the non-masked regions to mask code regions while blocking the opposite direction. Through this asymmetric information flow, the model can utilize the non-masked region without the mask code region negatively affecting the non-masked region. The biased self-attention is defined as follows: 
\begin{equation}
Attention(q, k, v) = \left(\text{SoftMax}\left(\frac{qk^T}{\sqrt{d}}\right)\odot \beta\right)v,
\end{equation}
where $\beta$ is the attention reweighting (AR) coefficient. Specifically, we define $\beta$ as 0.02 for mask positions and 1 for non-masked positions to achieve asymmetric information flow.

\subsection{Style direction adjustment}
\label{section:doubleblind}
After extracting the style, we adjust the directionality of the style to remove content information. To this end, we introduce two complementary losses as conceptually illustrated in Figure \ref{model} (c): style disentanglement (SD) loss encouraging orthogonality between style and content, and style preserving (SP) loss to align the style with prosody.
\subsubsection{Style disentanglement loss}
Inspired by \cite{10.1109/TASLP.2022.3164181}, we use orthogonality loss to disentangle style and content information. 
\begin{table*}[!ht]
    \centering
        \caption{Comparison with different models for subjective and objective metrics.}
    \label{Table1}\vspace{-0.3cm}
        \resizebox{0.95\textwidth}{!}{
    \begin{tabular}{l|c|cc|c|c|ccc|cc|c}
        \toprule
        \textbf{Method} & \textbf{Style Variation} &  \textbf{nMOS} ($\uparrow$) & \textbf{sMOS} ($\uparrow$) & \textbf{UTMOS} ($\uparrow$) &\textbf{WER} ($\downarrow$)  
         &$\textbf{RMSE}_{f0}$ ($\downarrow$) & $\textbf{RMSE}_{p}$ ($\downarrow$) & \textbf{F1 {V/UV}} ($\uparrow$) & \textbf{SECS} ($\uparrow$) \\ 
        \midrule
            GT & - & 4.34$\pm$0.04 & 4.63$\pm$0.02 & 3.78 & 12.42 & - & - & - & 0.9168 \\ 
            BigVGAN \cite{lee2023bigvgan}  &
            - &
            4.30$\pm$0.04 & 4.58$\pm$0.03 & 3.63 & 12.40 & 2.45 & 0.2749 & 0.8008 & 0.9150   \\ 
        \midrule
            FastSpeech 2 w/ GST \cite{pmlr-v80-wang18h}  & 
            \ding{55} &
            3.77$\pm$0.05 & 3.05$\pm$0.04 & 3.39 & 14.18 & 13.37 & 0.4619 & 0.6707 & 0.8945 \\
            FastSpeech 2 w/ CSE \cite{zhang-etal-2024-tcsinger} & \ding{51} 
             & 3.74$\pm$0.05 & 3.46$\pm$0.04 & 3.42 & 13.49 & 10.39 & 0.4159 & 0.7024 & 0.9013 \\
        \midrule
            StyleSpeech \cite{pmlr-v139-min21b}  & 
            \ding{55} &
            3.74$\pm$0.05 & 3.14$\pm$0.04 & 3.37 & 13.24 & 13.88 & 0.4433 & 0.6716 & 0.9008 \\ 
            GenerSpeech \cite{NEURIPS2022_4730d10b} & 
            \ding{51} &
            3.98$\pm$0.04 & 3.37$\pm$0.04  & 2.92 & 16.45 & 11.20 & 0.4343 & 0.6709 & 0.8848  \\
        \midrule
            Spotlight-TTS (Proposed)  & 
            \ding{51} & \textbf{4.26}$\pm$\textbf{0.04} & \textbf{3.84}$\pm$\textbf{0.04} & \textbf{3.56} & \ \textbf{12.64} & \textbf{8.27} & \textbf{0.4050} & \textbf{0.7053} & \textbf{0.9061} \\ 
        \bottomrule
    \end{tabular}
      }\vspace{-0.2cm}
\end{table*}
Since content information within style embedding can interfere with the learning of content embedding, we detach the content embedding during training. The SD loss $L_{\text{sd}}$ is defined as follows: 

\begin{equation}
L_{\text{sd}}
=  
\left\Vert
sg\bigl[E_{c}\bigr]\,
E_{s}^{T}
\right\Vert_{F}^{2},
\end{equation}
where \(E_{c}\) and \( E_{s}\) denote the content embedding, style embedding respectively.
The expression \(\left\Vert\cdot\right\Vert_{F}^{2}\) is the Frobenius norm.

\subsubsection{Style preserving loss}
To enhance stability in style learning and mitigate errors from F0 and V/UV predictions, we introduce an SP loss. This loss refines extracted style embeddings by increasing similarity with low-frequency prosody embeddings. Two separate MLP blocks are used to extract prosody embeddings from the lower 20 bins of the Mel-spectrogram and style embedding, respectively. By increasing the cosine similarity between these two embeddings, we refine the prosody information in the style embedding. The SP loss $L_{\text{sp}}$ is defined as follows: 
\begin{equation}
L_{\text{sp}} = -\sum_{i=1}^{t}\operatorname{cos\_sim}\bigl(p_i,\tilde{s}_i\bigr),
\end{equation}
where \(t\) is a number of timestep. \(p_i\) represents the prosody vector extracted from the low-band Mel-spectrogram, and \(\tilde{s}_i\) denotes the projected style vector. 

\subsection{Training stage of Spotlight-TTS}
As described in Figure \ref{model} (a), the proposed Spotlight-TTS model consists of a text encoder, a variance adaptor, a decoder, discriminators \cite{ijcai2022p620}, a style encoder, and MLP blocks. The MLP-blocks and discriminator are employed exclusively during the training phase.
For synthesizing the Mel-spectrogram, it is trained using the objective functions \(L_{\text{fs2}}\) of FastSpeech 2 \cite{ren2021fastspeech}.
 \( L_{\text{rvq}}\) is used to train the RVQ module.
As a result, the total loss of the proposed model is formulated as follows:
\begin{equation}
L_{\text{total}} = L_{\text{fs2}} + \lambda_{\text{rvq}} L_{\text{rvq}} + \lambda_{\text{adv}} L_{\text{adv}} + \lambda_{\text{sd}} L_{\text{sd}} + \lambda_{\text{sp}} L_{\text{sp}},
\end{equation}
where \( \lambda_{\text{rvq}}\), 
\( \lambda_{\text{adv}}\), \( \lambda_{\text{sd}}\), and \( \lambda_{\text{sp}}\) denotes the weights for each loss term, which are set to 1.0, 0.05, 0.02, and 0.02, respectively.

\section{Experiments and results}

\subsection{Experimental setup}
We use the emotional speech dataset (ESD) \cite{zhou2022emotional} to verify whether the models can capture style using expressive reference speech. It contains ten English speakers, each producing 350 sentences in five emotions (happy, sad, neutral, surprise, and angry). We follow the original partitioning criteria of the dataset, totaling 17,500 samples. Mel-spectrogram with 80 bins was extracted by short-time Fourier transform with a hop size of 256, a window size of 1,024, and an FFT size of 1,024. We employ the AdamW optimizer \cite{loshchilov2018decoupled}, setting the hyperparameters $\beta_{1}$ to 0.9 and $\beta_{2}$ to 0.98. Our model was trained for 200k steps on a single NVIDIA RTX 2080Ti GPU. For the audio synthesis in our experiments, we trained the vocoder using the official BigVGAN 
implementation \cite{lee2023bigvgan}. We compare our Spotlight-TTS with other FastSpeech2-based style transfer TTS models: FastSpeech2-GST \cite{ren2021fastspeech, pmlr-v80-wang18h}, FastSpeech2-CSE \cite{ren2021fastspeech,zhang-etal-2024-tcsinger}, StyleSpeech \cite{pmlr-v139-min21b}, GenerSpeech \cite{NEURIPS2022_4730d10b}. For models utilizing time-variant style, we use the same global style encoder \cite{NEURIPS2022_4730d10b} for sentence-level style.

\subsection{Implementation details}
Our TTS system is based on FastSpeech2 \cite{ren2021fastspeech} and incorporates a multi-length discriminator \cite{ijcai2022p620} for improved speech quality.
For SP loss, both style embedding and lower 20 bins of Mel-spectrogram are projected to 32 dimensions through MLP blocks from their initial dimensions of 256 and 20, respectively. Each MLP block consists of two linear layers with a GELU activation function \cite{hendrycks2016gaussian}. 
As shown in Figure \ref{model} (b), The style encoder is composed of two parts: a pre-trained global style encoder\footnote{https://github.com/Rongjiehuang/GenerSpeech/tree/encoder} that extracts the sentence-level style and a voiced-aware style encoder that focuses on frame-level style. The latter consists of a WaveNet block, four convolutional residual blocks, an RVQ module \cite{lee2022autoregressive}, and three UF blocks. The RVQ module has a depth of four and applies RT computed by a householder reflection matrix \cite{fifty2025restructuring}. After the unvoiced filler blocks, we use scaled dot-product attention \cite{NIPS2017_3f5ee243} to align the time dimension of the style embedding with that of the content embedding. The aligned style embedding is utilized within the variance adaptor and added to its final output along with the global style embedding. 

\subsection{Evaluation metrics}
To evaluate the speech quality, we conduct both objective and subjective evaluations of the synthesized speech. For the subjective evaluation, we use the naturalness mean opinion score (MOS) and similarity mean opinion score (sMOS). Both metrics are rated from 1 to 5 with a confidence interval of 95\%. We randomly select 50 samples from the test set, with 5 samples per speaker. Each audio has been listened to by
20 participants. Additionally, we utilize the open-source UTMOS\footnote{https://github.com/tarepan/SpeechMOS} \cite{saeki22c_interspeech} as a MOS prediction model for the naturalness metric. 
To evaluate linguistic consistency, we calculate the word error rate (WER) by Whisper\footnote{https://github.com/openai/whisper} \cite{pmlr-v202-radford23a}. 
For the speaker similarity measurements, we calculate the speaker embedding cosine similarity (SECS) via WavLM\footnote{https://huggingface.co/microsoft/wavlm-base-sv} \cite{chen2022wavlm}.
\begin{table}[!h]
\centering
\renewcommand{\arraystretch}{1.0}
\caption{Results of AXY preference test on parallel and non-parallel style transfer.}
\label{Table2} \vspace{-0.3cm}
\resizebox{1\columnwidth}{!}{
\begin{tabular}{l|c|c|ccc}
\toprule
\multirow[c]{2}{*}{\textbf{Baseline}} & \multirow[c]{2}{*}{\textbf{Setting}} & \multirow[c]{2}{*}{\textbf{7-point score}} & \multicolumn{3}{c}{\textbf{Preference (\%)}} \\
& & & X & Neutral & Y \\
\midrule
\multirow{2}{*}{FastSpeech2 w/ GST} & Parallel & $1.21\pm 0.12$ & 15\% & 22\% & 63\% \\
& Non-Parallel & $0.59\pm 0.11$ & 28\% & 16\% & 56\% \\
\midrule
\multirow{2}{*}{FastSpeech2 w/ CSE} & Parallel & $0.53\pm 0.12$ & 28\% & 17\% & 55\% \\
& Non-Parallel & $0.33\pm 0.15$ & 29\% & 29\% & 42\% \\
\midrule
\multirow{2}{*}{StyleSpeech} & Parallel & $1.16\pm 0.11$ & 17\% & 21\% & 62\% \\
& Non-Parallel & $0.25\pm 0.12$ & 34\% & 22\% & 44\% \\
\midrule
\multirow{2}{*}{GenerSpeech} & Parallel & $0.95\pm 0.12$ & 17\% & 36\% & 47\% \\
& Non-Parallel & $0.38\pm 0.13$ & 24\% & 33\% & 43\% \\
\bottomrule
\end{tabular}}
\vspace{-0.4cm}
\end{table}
For prosodic evaluation, we compute the root mean square error for both pitch error (RMSE$_{f_0}$) measured in Hz and periodicity error (RMSE$_{p}$), along with the F1 score of voiced/unvoiced classification (F1$_{v/uv}$). All objective metrics are computed using the entire test set. For reference audio selection, we used same strategy as in \cite{NEURIPS2022_4730d10b}.
\subsection{Model performance}
As shown in Table \ref{Table1}, our proposed Spotlight-TTS achieves significant improvements in both subjective and objective metrics. Global style-based models like FastSpeech2 w/GST and StyleSpeech demonstrate lower performance in style-related metrics due to the loss of temporal style variations through pooling operations. In contrast, FastSpeech2 w/CSE and GenerSpeech outperform global style-based models across all style-related metrics by preserving temporal dynamics. Furthermore, our model surpasses baselines across all metrics by considering temporal dynamics with region-specific importance and directionality of extracted style.

We also evaluate style transfer performance through an AXY test \cite{NEURIPS2022_4730d10b} with scores ranging from -3 to 3, where 0 denotes ``Both are about the same distance''. X, Neutral, Y denote baseline, same and ours respectively. As shown in Table \ref{Table2}, our subjective evaluation demonstrates superior style transfer performance across both parallel and non-parallel settings compared to baseline models. Similar to Table \ref{Table1}, time-variant style-based models outperform global style-based models.
Spotlight-TTS further advances this approach by explicitly modeling the relative importance of different temporal regions based on their acoustic characteristics. Unlike previous time-variant style-based models that treated all temporal regions equally, our region-aware style extraction enables more effective style capture, leading to superior style transfer performance. Additionally, our style direction adjustment disentangles content information from style in the embedding space, enabling robust style extraction even when linguistic content differs between reference speech and input text.

\subsection{Ablation study}
\subsubsection{Voiced-aware style extraction}
We evaluated voice-aware style extraction by removing key components and analyzing their effects. As shown in Table \ref{Table3}, removing RT leads to degraded performance, particularly in style-related metrics.
This indicates that the rotation trick allows the RVQ to better capture the harmonic structures in voiced regions by ensuring stable gradient propagation.
When both the RT and the UF are removed, we observe further performance degradation. Although unvoiced regions have relatively simple and repetitive patterns compared to voiced regions, removing the UF module leads to degraded pronunciation and prosody. 
This indicates that while unvoiced regions are less critical for style, proper handling of these regions through the UF module is still necessary.\begin{table}[h]
    \centering
    \caption{Results of ablation study on voiced-aware style extraction and style direction adjustment.}
    \label{Table3}\vspace{-0.3cm}
    \resizebox{1\columnwidth}{!}{
    \begin{tabular}{l|c|c|ccc}
        \toprule
        \textbf{Method} & \textbf{nMOS} & \textbf{WER} & $\textbf{RMSE}_{f0}$ & $\textbf{RMSE}_{p}$ & \textbf{F1 {V/UV}}\\ 
        \midrule
        Ours          & \textbf{3.93$\pm$0.07}         & \textbf{12.64}        & \textbf{8.27} & 
        0.4050 & \textbf{0.7053} \\ 
        \midrule
        $-$RT           & 3.91$\pm$0.06        & 13.24       & 9.43        & 0.4154  & 0.6928      \\ 
        $-$RT $-$UF       & 3.91$\pm$0.07       & 13.41       & 9.82  & 0.4274        & 0.6874          \\ 
        $-$RT $-$UF $-$VE   & 3.84$\pm$0.07        & 14.06       & 11.48  & 0.4425      & 0.6829          \\ 
        \midrule
        $-$SP           & 3.86$\pm$0.06          & 13.66       & 9.74     & 0.4297     & 0.6915        \\ 
        $-$SP $-$SD       & 3.66$\pm$0.07          & 15.38       & 8.53    & \textbf{0.4037}      & 0.6848  \\
        \bottomrule
    \end{tabular}
    }\vspace{-0.4cm}
\end{table}
\begin{table}[h]
    \centering
    \caption{Results of ablation study on biased self-attention in unvoiced filler.}
    \label{Table4}\vspace{-0.3cm}
    \resizebox{1\columnwidth}{!}{
    \begin{tabular}{l|c|c|ccc}
        \toprule
        \textbf{Method} &  \textbf{nMOS} &
        \textbf{WER}  &
        $\textbf{RMSE}_{f0}$ & $\textbf{RMSE}_{p}$ & \textbf{F1 {V/UV}}\\ 
        \midrule
        Ours &  \textbf{3.95$\pm$0.07}     
        & \textbf{12.64}
        & \textbf{8.27} & \textbf{0.4050} & \textbf{0.7053}  \\ 
        \midrule
        Self-attention w/ BM          &  3.93$\pm$0.07
        & 13.61
        & 13.19    & 0.4528  & 0.6751     \\ 
        Self-attention  & 3.87$\pm$0.07  &  
        13.64
        & 16.38  & 0.4587   & 0.6668      \\ 
        \bottomrule
    \end{tabular}
    }\vspace{-0.4cm}
\end{table} Additionally, removing VE results in significantly increased pitch error. This demonstrates that our voiced-aware approach, which aggregates voiced frames to focus quantization on style-rich regions, is essential for capturing detailed style information. 
\subsubsection{Style direction adjustment}We also investigate the effectiveness of our style direction adjustment mechanism. When the SP loss is removed, we observe highly increased pitch errors. 
This suggests that SP loss effectively mitigates the degradation of prosodic information caused by the strong constraints of SD loss. Removing both SP and SD losses results in severely degraded nMOS and WER indicating that removing content information from style is important for both speech quality and pronunciation.
\subsubsection{Biased self-attention}
To investigate the effectiveness of biased self-attention, we conduct an additional ablation in Table~\ref{Table4}. If we replace AR with a simple binary mask (BM) with 1 and 0, the model completely blocks information flow from non-masked to mask code regions, preventing the mask code embeddings from being filled with meaningful acoustic information. This disrupts the natural prosodic continuity between non-masked and mask code regions, resulting in higher F0 errors and worse V/UV classification. In contrast, using conventional self-attention allows excessive interference between regions, significantly degrading both metrics. These results validate that our biased self-attention enables optimal information flow between voiced and unvoiced regions.

\section{Conclusion}
We presented Spotlight-TTS, a framework for synthesizing expressive speech by focusing on voiced regions in the Mel-spectrogram and adjusting the direction of the extracted style. Voiced-aware style extraction considers the acoustic characteristics of different speech regions, enabling more detailed style extraction. Furthermore, the style direction adjustment effectively disentangles content from style, while preserving prosody information within the style embedding. Experimental results demonstrate that our method generates more natural, expressive speech while achieving these improvements through style-focused modifications. Despite these advances, there remains room for improvement in non-parallel style transfer, especially when reference speech duration significantly differs from the input text. Future work will focus on improving speech quality in non-parallel settings. Nevertheless, our findings highlight the effectiveness of sophisticated style modeling, offering a promising direction for expressive TTS systems. 

\section{Acknowledgements}
This work was partly supported by the Institute of Information \& Communications Technology Planning \& Evaluation (IITP) grant funded by the Korea government (MSIT) (Artificial Intelligence Graduate School Program (Korea University) (No. RS-2019-II190079), Artificial Intelligence Innovation Hub (No. RS-2021-II212068), AI Technology for Interactive Communication of Language Impaired Individuals (No. RS-2024-00336673), and Artificial Intelligence Star Fellowship Support Program to Nurture the Best Talents (IITP-2025-RS-2025-02304828)).

\bibliographystyle{IEEEtran}
\bibliography{mybib}

\begin{thebibliography}{10}
\providecommand{\url}[1]{#1}
\csname url@samestyle\endcsname
\providecommand{\newblock}{\relax}
\providecommand{\bibinfo}[2]{#2}
\providecommand{\BIBentrySTDinterwordspacing}{\spaceskip=0pt\relax}
\providecommand{\BIBentryALTinterwordstretchfactor}{4}
\providecommand{\BIBentryALTinterwordspacing}{\spaceskip=\fontdimen2\font plus
\BIBentryALTinterwordstretchfactor\fontdimen3\font minus \fontdimen4\font\relax}
\providecommand{\BIBforeignlanguage}[2]{{%
\expandafter\ifx\csname l@#1\endcsname\relax
\typeout{** WARNING: IEEEtran.bst: No hyphenation pattern has been}%
\typeout{** loaded for the language `#1'. Using the pattern for}%
\typeout{** the default language instead.}%
\else
\language=\csname l@#1\endcsname
\fi
#2}}
\providecommand{\BIBdecl}{\relax}
\BIBdecl

\bibitem{ren2021fastspeech}
Y.~Ren, C.~Hu, X.~Tan, T.~Qin, S.~Zhao, Z.~Zhao, and T.-Y. Liu, ``Fastspeech 2: Fast and high-quality end-to-end text to speech,'' in \emph{International Conference on Learning Representations}, 2021.

\bibitem{557671}
S.-W. Lee and H.-H. Song, ``A new recurrent neural-network architecture for visual pattern recognition,'' \emph{IEEE Transactions on Neural Networks}, vol.~8, no.~2, pp. 331--340, 1997.

\bibitem{602073}
S.-W. Lee and Y.-J. Kim, ``Multiresolution recognition of handwritten numerals with wavelet transform and multilayer cluster neural network,'' in \emph{Proceedings of 3rd International Conference on Document Analysis and Recognition}, vol.~2, 1995, pp. 1010--1013 vol.2.

\bibitem{LEE1995783}
S.-W. Lee, ``Multilayer cluster neural network for totally unconstrained handwritten numeral recognition,'' \emph{Neural Networks}, vol.~8, no.~5, pp. 783--792, 1995.

\bibitem{NEURIPS2022_69c754f5}
S.-H. Lee, S.-B. Kim, J.-H. Lee, E.~Song, M.-J. Hwang, and S.-W. Lee, ``Hierspeech: Bridging the gap between text and speech by hierarchical variational inference using self-supervised representations for speech synthesis,'' in \emph{Advances in Neural Information Processing Systems}, vol.~35, 2022, pp. 16\,624--16\,636.

\bibitem{cho24_interspeech}
D.-H. Cho, H.-S. Oh, S.-B. Kim, S.-H. Lee, and S.-W. Lee, ``Emosphere-tts: Emotional style and intensity modeling via spherical emotion vector for controllable emotional text-to-speech,'' in \emph{Interspeech 2024}, 2024, pp. 1810--1814.

\bibitem{pmlr-v80-skerry-ryan18a}
R.~Skerry-Ryan, E.~Battenberg, Y.~Xiao, Y.~Wang, D.~Stanton, J.~Shor, R.~Weiss, R.~Clark, and R.~A. Saurous, ``Towards end-to-end prosody transfer for expressive speech synthesis with tacotron,'' in \emph{International Conference on Machine Learning}, vol.~80, 2018, pp. 4693--4702.

\bibitem{pmlr-v80-wang18h}
Y.~Wang, D.~Stanton, Y.~Zhang, R.-S. Ryan, E.~Battenberg, J.~Shor, Y.~Xiao, Y.~Jia, F.~Ren, and R.~A. Saurous, ``Style tokens: Unsupervised style modeling, control and transfer in end-to-end speech synthesis,'' in \emph{International Conference on Machine Learning}, vol.~80, 2018, pp. 5180--5189.

\bibitem{pmlr-v139-min21b}
D.~Min, D.~B. Lee, E.~Yang, and S.~J. Hwang, ``Meta-stylespeech : Multi-speaker adaptive text-to-speech generation,'' in \emph{International Conference on Machine Learning}, vol. 139, 2021, pp. 7748--7759.

\bibitem{tang2023qi}
H.~Tang, X.~Zhang, J.~Wang, N.~Cheng, and J.~Xiao, ``Qi-tts: Questioning intonation control for emotional speech synthesis,'' in \emph{2023 IEEE International Conference on Acoustics, Speech and Signal Processing}, 2023, pp. 1--5.

\bibitem{NEURIPS2022_4730d10b}
R.~Huang, Y.~Ren, J.~Liu, C.~Cui, and Z.~Zhao, ``Generspeech: Towards style transfer for generalizable out-of-domain text-to-speech,'' in \emph{Advances in Neural Information Processing Systems}, vol.~35, 2022, pp. 10\,970--10\,983.

\bibitem{NIPS2017_7a98af17}
A.~van~den Oord, O.~Vinyals, and k.~kavukcuoglu, ``Neural discrete representation learning,'' in \emph{Advances in Neural Information Processing Systems}, vol.~30, 2017.

\bibitem{seong24b_interspeech}
D.~Seong, H.~Lee, and J.-H. Chang, ``Tsp-tts: Text-based style predictor with residual vector quantization for expressive text-to-speech,'' in \emph{Interspeech 2024}, 2024, pp. 1780--1784.

\bibitem{zhang-etal-2024-tcsinger}
Y.~Zhang, Z.~Jiang, R.~Li, C.~Pan, J.~He, R.~Huang, C.~Wang, and Z.~Zhao, ``{TCS}inger: Zero-shot singing voice synthesis with style transfer and multi-level style control,'' in \emph{Proceedings of the 2024 Conference on Empirical Methods in Natural Language Processing}, 2024, pp. 1960--1975.

\bibitem{bengio2013estimating}
Y.~Bengio, N.~L{\'e}onard, and A.~Courville, ``Estimating or propagating gradients through stochastic neurons for conditional computation,'' \emph{arXiv preprint arXiv:1308.3432}, 2013.

\bibitem{jayant1993signal}
N.~Jayant, J.~Johnston, and R.~S. Safranek, ``Signal compression based on models of human perception,'' \emph{Proceedings of the IEEE}, vol.~81, no.~10, pp. 1385--1422, 1993.

\bibitem{huang2023not}
M.~Huang, Z.~Mao, Q.~Wang, and Y.~Zhang, ``Not all image regions matter: Masked vector quantization for autoregressive image generation,'' in \emph{Proceedings of the IEEE/CVF Conference on Computer Vision and Pattern Recognition}, 2023, pp. 2002--2011.

\bibitem{Liu_Gui_Luo_2023}
Z.~Liu, J.~Gui, and H.~Luo, ``Good helper is around you: Attention-driven masked image modeling,'' \emph{Proceedings of the AAAI Conference on Artificial Intelligence}, vol.~37, no.~2, pp. 1799--1807, 2023.

\bibitem{fifty2025restructuring}
C.~Fifty, R.~G. Junkins, D.~Duan, A.~Iyengar, J.~W. Liu, E.~Amid, S.~Thrun, and C.~Re, ``Restructuring vector quantization with the rotation trick,'' in \emph{International Conference on Learning Representations}, 2025.

\bibitem{lee2022autoregressive}
D.~Lee, C.~Kim, S.~Kim, M.~Cho, and W.-S. Han, ``Autoregressive image generation using residual quantization,'' in \emph{Proceedings of the IEEE/CVF Conference on Computer Vision and Pattern Recognition}, 2022, pp. 11\,523--11\,532.

\bibitem{liu2022convnet}
Z.~Liu, H.~Mao, C.-Y. Wu, C.~Feichtenhofer, T.~Darrell, and S.~Xie, ``A convnet for the 2020s,'' in \emph{Proceedings of the IEEE/CVF Conference on Computer Vision and Pattern Recognition}, 2022, pp. 11\,976--11\,986.

\bibitem{10.1109/TASLP.2022.3164181}
T.~Li, X.~Wang, Q.~Xie, Z.~Wang, and L.~Xie, ``Cross-speaker emotion disentangling and transfer for end-to-end speech synthesis,'' \emph{IEEE/ACM Trans. Audio, Speech and Lang. Proc.}, vol.~30, p. 1448–1460, 2022.

\bibitem{lee2023bigvgan}
S.-g. Lee, W.~Ping, B.~Ginsburg, B.~Catanzaro, and S.~Yoon, ``Bigvgan: A universal neural vocoder with large-scale training,'' in \emph{International Conference on Learning Representations}, 2023.

\bibitem{ijcai2022p620}
Z.~Ye, Z.~Zhao, Y.~Ren, and F.~Wu, ``Syntaspeech: Syntax-aware generative adversarial text-to-speech,'' in \emph{Proceedings of the Thirty-First International Joint Conference on Artificial Intelligence, {IJCAI-22}}, 2022, pp. 4468--4474, main Track.

\bibitem{zhou2022emotional}
K.~Zhou, B.~Sisman, R.~Liu, and H.~Li, ``Emotional voice conversion: Theory, databases and esd,'' \emph{Speech Communication}, vol. 137, pp. 1--18, 2022.

\bibitem{loshchilov2018decoupled}
I.~Loshchilov and F.~Hutter, ``Decoupled weight decay regularization,'' in \emph{International Conference on Learning Representations}, 2019.

\bibitem{hendrycks2016gaussian}
D.~Hendrycks and K.~Gimpel, ``Gaussian error linear units (gelus),'' \emph{arXiv preprint arXiv:1606.08415}, 2016.

\bibitem{NIPS2017_3f5ee243}
A.~Vaswani, N.~Shazeer, N.~Parmar, J.~Uszkoreit, L.~Jones, A.~N. Gomez, L.~u. Kaiser, and I.~Polosukhin, ``Attention is all you need,'' in \emph{Advances in Neural Information Processing Systems}, vol.~30, 2017.

\bibitem{saeki22c_interspeech}
T.~Saeki, D.~Xin, W.~Nakata, T.~Koriyama, S.~Takamichi, and H.~Saruwatari, ``Utmos: Utokyo-sarulab system for voicemos challenge 2022,'' in \emph{Interspeech 2022}, 2022, pp. 4521--4525.

\bibitem{pmlr-v202-radford23a}
A.~Radford, J.~W. Kim, T.~Xu, G.~Brockman, C.~Mcleavey, and I.~Sutskever, ``Robust speech recognition via large-scale weak supervision,'' in \emph{International Conference on Machine Learning}, vol. 202, 2023, pp. 28\,492--28\,518.

\bibitem{chen2022wavlm}
S.~Chen, C.~Wang, Z.~Chen, Y.~Wu, S.~Liu, Z.~Chen, J.~Li, N.~Kanda, T.~Yoshioka, X.~Xiao \emph{et~al.}, ``Wavlm: Large-scale self-supervised pre-training for full stack speech processing,'' \emph{IEEE Journal of Selected Topics in Signal Processing}, vol.~16, no.~6, pp. 1505--1518, 2022.

\end{thebibliography}

\end{document}